# Classifying variable-structures: a general framework

Xavier Bry [a], Lionel Cucala [a]

[a] Institut Montpelliérain Alexander Grothendieck, Université de Montpellier,
Place Eugène Bataillon CC 051 - 34095, Montpellier, France

*Abstract:*

*In this work, we unify recent variable-clustering techniques within a common geometric framework which allows to extend clustering to variable-structures, i.e. variable-subsets within which links between variables are taken into consideration in a given way. All variables being measured on the same $n$ statistical units, we first represent every variable-structure with a unit-norm operator in $\mathbb{R}^{n \times n}$. We consider either the euclidean chord-distance or the geodesic distance on the unit-sphere of $\mathbb{R}^{n \times n}$. Then, we introduce the notion of rank-H average of such operators as the rank-H solution of a compound distance-minimisation program. Finally, we propose a K-means-type algorithm using the rank-H average as centroid to perform variable-structure clustering. The method is tested on simulated data and applied to wine data.*

## 1. Introduction

### 1.1. Framework and goal

Consider a data-array describing $n$ statistical units through $p$ variables which may be of different types, i.e. numerical or categorical. High-dimensional data, i.e. where $p > n$, call for dimension reduction. One way to perform this in a non-supervised way is to cluster variables and replace each cluster with a lesser number of appropriate dimensions which can aptly stand for the cluster in subsequent models of the data.

In a multi-array setting, where units are described through many subsets of variables, each subset might be viewed as a separate piece of information, and one may subsequently want to perform clustering on these subsets.

### 1.2. A quick review of existing methods

[Escoufier 1970] considered that the $p$ variables could be viewed as a sample taken from a population of numerical variables, just as the units were from a population of units. Taking it that the affine scaling of a numerical variable being arbitrary and irrelevant to informational issues, he proposed to standardise every such variable, hence to represent it with a vector on the unit sphere $S_{n-1} \subset \mathbb{R}^n$, on which he defined a theoretical framework for sampling numerical variables. In his wake, [Gomes P. 1987] and [Gomes A. 1993] addressed the problem of partitioning a set of $p$ numerical variables. This set was modelled through a mixture of $k$ Bingham distributions, and the parameters of the mixture were estimated through a K-means-type algorithm. Interesting theoretical results were produced regarding estimation performance. The stability of the algorithm's fixed point was also studied. A first limitation of the method is that it tackles numerical variables only. A second limitation of this parametric approach is the constraint that each component of the mixture is Bingham-distributed. An advantage of using Bingham-distributed mixture-components is that the theoretical mean of each component is its mode, which is one of the parameters being estimated. The third limitation is that the algorithm assumes $k$ known. Thus, several values of $k$ are to be tried, and the subsequent partitions compared with respect to some criterion.

More recently, [Qannari et al. 1998] have proposed to represent every variable, be it numerical or categorical (then coded by its centred indicator-variables less one to avoid multicollinearity), by the orthogonal projector on the subspace of $\mathbb{R}^n$ associated with it. They then propose to endow the space $\mathbb{R}^{n \times n}$ where these projectors exist with the Frobenius scalar product. This scalar product between projectors, and the associated cosine, give back classical statistical measures. This euclidean representation in an operator-space allows a very satisfactory unification of the distance between variables, and this is the framework we shall be working from. Finally, the authors use this distance for variable-clustering in the K-means spirit. K-means entail the definition of a centroid for each cluster. The centroid they use is the average of the projectors in the cluster. We shall question this choice in section 2.5.

A few years later, [Vigneau et al. 2003, 2006] have proposed to cluster numerical variables about estimated numerical latent variables. They classically represent the variables as vectors in $\mathbb{R}^n$. Their initial method uses a K-means-type algorithm iterating a classical factor-model estimation in each cluster (yielding the 1$^{st}$ principal component of the cluster as "centroid"). The method is then extended to latent variables pertaining to a space spanned by external variables, and is linked to Instrumental Variables Principal Component Analysis (PCA) and multivariate partial least squares (PLS) regression. Finally, [Saracco et al. 2010, Chavent et al. 2010, 2012] extended the previous method to categorical variables.

In this paper, we show that the methods by [Qannari et al. 1998, Vigneau et al. 2003, Saracco et al. 2010, Chavent et al. 2010, 2012] are very closely related, and can be unified within the mathematical framework of the unit sphere in $\mathbb{R}^{n \times n}$ proposed by [Qannari et al. 1998]. We will show how any variable-subset can be represented by a corresponding operator on this sphere. This framework allows to extend the above-mentioned methods in four respects:

1. Not only variables will be able to be clustered, but more generally variable-structures, each based on a particular variable-subset.
2. The use of a Reproducing Kernel Hilbert Space (RKHS) will allow the data-structures to go beyond the linear potential of variables.
3. The structure of each cluster will be used beyond its first principal component (PC) in the clustering process.
4. We also propose, as an alternative to the euclidean distance, to use the geodesic distance on the unit sphere.

*1.5. Plan of the paper*

In section 2, we give the geometric representation of the data and problem. Section 3 gives a first basic approach to variable-clustering. In section 4, we present our extension. Section 5 benchmarks the method on simulated data and finally applies it to wine-data.

## 2. Geometric representations of the data

Each unit $i$ is endowed with a weight $w_i > 0$. We suppose $\sum_{i=1}^{n} w_i = 1$ and denote: $W := diag(w_i; i=1,...,n) \in \mathbf{M}_n$, where $\mathbf{M}_n$ is the set of $n \times n$ matrices. Additional notations can be found in appendix 1.

## 2.1. The Variable-Space

**Numerical variables**

Numerical variable $x$ is identified with vector $x \in \mathbb{R}^n$. We will refer to $\mathbb{R}^n$ as the Variable-Space (V-Space). It is endowed with scalar product :

$$\langle x|y\rangle_W = x'Wy, \forall x,y \in \mathbb{R}^n$$

All variables are taken centred, so that, **1** being the vector having all values equal to 1:

$$\forall x : \langle x|\mathbf{1}\rangle_W = 0$$

$$\forall x,y : \langle x|y\rangle_W = cov(x;y) \; ; \; \|x\|_W^2 = v(x)$$

Let $S_{n-1}$ denote the unit-radius sphere in $\mathbb{R}^n$, and:

$$S_V := S_{n-1} \cap \langle \mathbf{1} \rangle^\perp$$

Standardized variables are identified with vectors of $S_V$.

**Categorical variables**

A categorical variable having $r$ levels is primarily coded through the corresponding $r$ indicator-variables. But to concentrate on the non-redundant information and avoid multicollinearity, one of the indicator-variables is removed and the remaining ones are centred. We shall identify the variable with the matrix $X$ whose columns are the retained $r-1$ centred indicator variables. Thus, categorical variables cannot be represented as one vector in the V-space. This will motivate a change of space, in section 2.3.

## 2.2. Concentrating on directions

**The need for a metric on directions**

Let $x \in S_V$ ; $x$ and $-x$ being poles apart on $S_V$, the distance between them is maximum (= 2), though they are equivalent from the informational point of view. We hereby see that, should we endow $S_V$ with a metric, this could not just be that induced by the euclidean distance associated with the above-given scalar product. We clearly need a metric $\delta$ which is such that $\delta(x;x)=0$, and that $\delta(x;y)$ depends only on the angle between $\langle x \rangle$ and $\langle y \rangle$, so that $\delta(x;-x)=0$. Such a metric can always be given the following general form:

$$\delta(x,y) = g(\langle x|y\rangle^2) \quad (1),$$

where $g$ is non-negative, decreasing, and $g(1)=0$

Let us now introduce categorical variables, and formalise the operator-based geometric approach to classical bivariate statistical relationships proposed by [Qannari et al. 1998].

**Bivariate statistical relationships**

The square correlation between two standardized variables $x$ and $y$ is:

$$\langle x|y\rangle_W^2 = x'Wy\,y'Wx = tr(x'Wy\,y'Wx) = tr(xx'W\,yy'W) = tr(\Pi_x^* \Pi_y) \text{ , where}$$

$\forall A(n,n), A^* = W^{-1}A'W$ is the adjoint of $A$ with respect to the $W$-scalar product.

The analysis-of-variance ratio $R^2$ between a standardized numerical variable $y$ and a categorical one $X$ is:

$$R^2(y,X) = \cos^2(y,\langle X \rangle) = \langle y|\Pi_X y\rangle W = y'W\Pi_X y = tr(yy'W\Pi_X) = tr(W^{-1}\Pi_y'W\Pi_X)$$

Finally, if $X$ and $Y$ are categorical variables, the $\Phi^2(X,Y) = \frac{1}{n}\chi^2(X,Y)$ coefficient measuring the intensity of their relationship is (cf appendix 2-a):
$$\Phi^2(X,Y) = tr(\Pi_Y^* \Pi_X)$$
This unified expression of bivariate relationship intensities incites us to consider representing each variable through the projector on its subspace of the V-space. Such projectors belong to another space: the operator-space.

## 2.3. The Operator-Space

$\forall m \in \mathbb{N}, \forall X \in \mathbb{R}^{n \times m}$, subspace $\langle X \rangle$ may be uniquely represented by the orthogonal projector on it: $\Pi_X = X(X'WX)^{-1} X'W \in \mathbb{R}^{n \times n}$. We shall refer to $\mathbb{R}^{n \times n}$ as the Operator-space (O-space). It can be endowed with the following scalar product:
$$\forall A, B \in \mathbb{R}^{n \times n} : [A|B] := tr(A^* B) \quad (2)$$
The unit-sphere of the O-space will be denoted $S_O$.

**Orthogonal projectors**

If $dim(\langle X \rangle) = r$:
$$\llbracket \Pi_X \rrbracket^2 := [\Pi_X | \Pi_X] = tr(\Pi_X^* \Pi_X) = tr(\Pi_X^2) = tr(\Pi_X) = r$$
We can thus norm projectors. For every numerical variable, the associated projector is already unit-norm. For a categorical variable $X$ with $m$ levels:
$$dim(\langle X \rangle) = m - 1 \Rightarrow \llbracket \Pi_X \rrbracket = \sqrt{m-1}$$
Let $\tilde{X} := \dfrac{\Pi_X}{\llbracket \Pi_X \rrbracket}$. Note that for two categorical variables $X$ and $Y$ with respectively $r$ and $s$ levels:
$$\cos^2(\Pi_X, \Pi_Y) = [\tilde{X} | \tilde{Y}] = \frac{\Phi^2(X,Y)}{\sqrt{r-1}\sqrt{s-1}} := T(X,Y),$$
where $T^2(X,Y)$ is the Tschuprow coefficient of the two variables.

Thus, with every variable we can associate a unique vector of $S_O$: the normed orthogonal projector on the corresponding subspace of $\mathbb{R}^n$, which only takes into account the directional information of the variable, and allows to formulate the classical bivariate statistical links.

**Beyond orthogonal projectors: resultants**

Let $X$ now be any $(n \times q)$ matrix, the columns of which code a set of variables of any types. Let $M$ be a ($q \times q$) matrix, such that PCA of $X$ with metric $M$ and weights $W$ can be given a relevant interpretation. Informally, $X$ codes the data, and $M$ codes the way the data is to be looked at, i.e. the purpose of $M$ is to reveal or conceal some correlation structures within $X$. [Bry 2001] has defined and termed *resultant* of $X$ with respect to metric $M$ (and weights $W$) the following operator:
$$R_{X,M} := XMX'W$$
Note that this operator is also the inertia operator used in the PCA of $(X, M, W)$. Projectors are mere particular cases: $\forall X, \Pi_X = R_{X,(X'WX)^{-1}}$. In their case, metric $M = (X'WX)^{-1}$ has the effect of erasing any correlation structure within $X$, only leaving its subspace-related information.

**Properties of the resultant-set**

1) The set $\mathbf{R}_n$ of $n \times n$ $W$-resultants is exactly that of $W$-symmetric positive

semi-definite (spsd) operators. Indeed, every resultant is $W$-spsd, and conversely: let $A$ be any $W$-spsd operator in $S_O$. Diagonalisation of $A$ allows to write:
$$A = U \Lambda U' W$$
where $U$ is an orthogonal matrix, and $\Lambda$ a diagonal one with non-negative diagonal elements.

As a consequence, $\forall A, B \in \mathbf{R}_n : A^* = A \Rightarrow [A|B] = tr(AB)$.

2) $\mathbf{R}_n$ is stable under linear combination with non-negative weights. Thus, it is a convex cone. As a consequence, the weighted average of $W$-spsd operators is one.

**r-equivalence**

We shall from now on consider the normed resultant and denote it:
$$\tilde{R}_{X,M} := r(X, M) := \frac{R_{X,M}}{[\![ R_{X,M} ]\!]}$$

To sum things up, to every ordered pair $(X, M)$, where $X$ is a $(n \times q)$ data matrix coding a set of variables, and $M$ a $(q \times q)$ metric matrix, we can associate a unique vector of $S_O$ : $r(X, M)$.

<u>Definition</u>: Two ordered pairs $(X, M)$ and $(Y, N)$ will be said to be *r-equivalent* (denoted $(X, M) \stackrel{r}{\sim} (Y, N)$ ) iff:
$$r(X, M) = r(Y, N)$$

$\forall X, M$ and $\Omega$ orthogonal matrix, all with suitable size, we obviously have:
$$(XM^{1/2}\Omega, I) \stackrel{r}{\sim} (X, M)$$

Besides, mapping $r$ is 0-degree homogenous with respect to either of its arguments:
$$\forall \alpha \neq 0, \beta > 0, r(\alpha X, \beta M) = r(X, M) \quad (3)$$

As a consequence, a given normed $W$-spsd operator in $S_O$ can represent an infinity of variable sets.

**Scalar products of resultants**

$\forall X(n, p), Y(n, q)$, and $M(p, p), N(q, q)$ spd:
$$[R_{X,M} | R_{Y,N}] = tr(XMX'WYNY'W) \quad (4)$$
$$= tr(M^{1/2}X'WYNY'WXM^{1/2}) = tr((\tilde{X}'W\tilde{Y})(\tilde{X}'W\tilde{Y})') \text{, with: } \tilde{X} = XM^{1/2}, \tilde{Y} = YN^{1/2}$$
$$= \sum_{\substack{j=1,p \\ k=1,q}} \langle \tilde{x}^j | \tilde{y}^k \rangle_W^2 \quad (5)$$

This leads to interesting statistical interpretations according to the case.

Example 1:
> If the columns $x^j$ of $X$ and $y^k$ of $Y$ are centred numerical variables, and: $M = diag(1/\sigma^2(x^j))_j$, $N = diag(1/\sigma^2(y^k))_k$, then the columns $\tilde{x}^j$ of $\tilde{X}$ and $\tilde{y}^k$ of $\tilde{Y}$ are the corresponding standardised variables, and:
> $$[R_{X,M} | R_{Y,N}] = \sum_{\substack{j=1,p \\ k=1,q}} \rho^2(x^j, y^k)$$

Example 2:
> If the columns $x^j$ of $X$ are centred numerical variables, and those of $Y$ are the $q$ indicator-variables of a categorical variable, then, with

$$M = diag(1/\sigma^2(x^j))_j \text{ and } N = (Y'WY)^{-1} :$$
$$(4) \Rightarrow [R_{X,M}|R_{Y,N}] = tr(\tilde{X}\tilde{X}'W\Pi_Y) = tr(\tilde{X}\tilde{X}'W\Pi_Y^2)$$
$$= tr(\tilde{X}'W\Pi_Y^2\tilde{X}) = tr(\tilde{X}_Y'W\tilde{X}_Y)$$

where $\tilde{X}_Y = \Pi_Y \tilde{X}$ is the matrix having as columns the regressions of the $x^j$'s on $Y$, i.e. containing the means of the $x^j$'s conditional on $Y$. As a consequence, $[R_{X,M}|R_{Y,N}]$ is the between-$Y$-group inertia of the units coded in $X$.

**Angle between resultants**

From (2), we can derive the square cosine between two resultants:

$$\cos^2(\tilde{R}_{X,M},\tilde{R}_{Y,N}) = [\tilde{R}_{X,M}|\tilde{R}_{Y,N}]^2 = \frac{[R_{X,M}|R_{Y,N}]}{[\![R_{X,M}]\!][\![R_{Y,N}]\!]}$$

This is no other than Escoufier's RV-coefficient [Robert et al. 1976].

### *2.4. Distances in the Operator-space*

#### 2.4.1. The euclidean distance

(2) defines a norm, hence a euclidean distance in $\mathbb{R}^{n \times n}$, which we can use.

The square distance between two points $\tilde{A}$ and $\tilde{B}$ of $S_O$ is then:

$$d^2(\tilde{A},\tilde{B}) = [\![\tilde{A}-\tilde{B}]\!]^2 = [\tilde{A}-\tilde{B}|\tilde{A}-\tilde{B}] = [\![\tilde{A}]\!]^2 + [\![\tilde{B}]\!]^2 - 2[\tilde{A}|\tilde{B}] = 2(1-[\tilde{A}|\tilde{B}]) \quad (6)$$

For a pair of standardised variables $x$ and $y$, for instance, we get:

$$[\![xx'W - yy'W]\!]^2 = 2(1 - cos(xx'W, yy'W)) = 2(1 - \langle x|y\rangle^2)$$

We shall refer to this euclidean distance on $S_O$ as the "chord-distance".

#### 2.4.2. The geodesic distance on $S_O$

Alternatively, we can look upon $S_O$ as a variety in the O-space and use the geodesic distance to measure discrepancies on it. The geodesic distance between two points $\tilde{A}$ and $\tilde{B}$ of $S_O$ is, up to a multiplicative factor, the length of their arc, i.e.:

$$\delta(\tilde{A},\tilde{B}) = \cos^{-1}([\tilde{A}|\tilde{B}]) \quad (7)$$

## 3. Averaging resultants

### *3.1. Averaging resultants using the euclidean distance d*

Here, we are tackling the definition of a suitable weighted average of a system of normed resultants $\{\tilde{R}_1,...,\tilde{R}_K\} \subset S_O$ with respect to a weight-system $\Omega = \{\omega_1,...,\omega_K\}, \sum_k \omega_k = 1$.

#### 3.1.1. The classical weighted average

The classical weighted average is:

$$\bar{R}_\Omega = \arg\min_{R \in \mathbf{M}_n} \sum_{k=1}^K \omega_k [\![\tilde{R}_k - R]\!]^2 = \sum_k \omega_k \tilde{R}_k \quad ,$$

And we also have: $\bar{R}_\Omega \in \mathbf{R}_n \Rightarrow \bar{R}_\Omega = \arg\min_{R \in \mathbf{R}_n} \sum_{k=1}^K \omega_k [\![\tilde{R}_k - R]\!]^2$

The sphere being convex, this average belongs to the closed ball, and to $S_O$ if and only if all $\tilde{R}_k$'s having non-zero weight are equal. But we are interested in an average constrained to belong to $S_O$.

### 3.1.2. The $S_O$-constrained average

The Huygens theorem yields:

$$\forall R \in S_O : \sum_{k=1}^{K} \omega_k \|\tilde{R}_k - R\|^2 = \sum_{k=1}^{K} \omega_k \|\tilde{R}_k - \bar{R}\|^2 + \|R - \bar{R}\|^2 \quad (8)$$

So:

$$\arg\min_{R \in \mathbf{R}_n, \|R\|=1} \sum_{k=1}^{K} \omega_k \|\tilde{R}_k - R\|^2 = \arg\min_{R \in \mathbf{R}_n, \|R\|=1} \|R - \bar{R}\|^2 = \frac{\bar{R}}{\|\bar{R}\|} \quad (9)$$

Indeed, the projection on the sphere of any point in the ball other than its centre is the extremity of the corresponding radius.

Now, when each $\tilde{R}_k$ represents a data-array $X_k$, we can easily exhibit a particular compound data-array which the normed weighted average represents:

$\forall (X, M)$, with $X(n \times q)$, we define $X^* := XM^{1/2}$. We have: $R_{X,M} = R_{X^*, I_p}$.

Now $\forall X(n \times q)$, $\forall Y(n \times s)$, $\forall M(q \times q), N(s \times s)$ sdp and $\forall \alpha, \beta > 0$:

$$\alpha R_{X,M} + \beta R_{Y,N} = R_{[\sqrt{\alpha} X^*, \sqrt{\beta} Y^*], I_{p+q}} \quad (10)$$

As a consequence, $\forall \{\omega_h \geq 0; h = 1, H\}$ s.t. $\sum_{h=1}^{H} \omega_h = 1$, $\bar{R} = \sum_{h=1}^{H} \omega_h R_{X_h, M_h}$ is such that:

$$\frac{\bar{R}}{\|\bar{R}\|} = r([\sqrt{\omega_1} X_1^*, \dots, \sqrt{\omega_H} X_H^*], I) \quad (11),$$

which represents on $S_O$ the data-matrix: $[\sqrt{\omega_1} X_1^*, \dots, \sqrt{\omega_H} X_H^*]$.

### 3.1.3. The euclidean rank-$H$ average

The previous average may contain too many dimensions (including noise) to be that interesting. We shall now look for an average constrained not only to be on $S_O$, but to have rank $H$, where $H$ is lower or equal to $rank(\bar{R})$. The set of all rank-$H$ resultants is obviously a cone $C_H$. So, we are interested in finding:

$$\arg\min_{R \in C_H \cap S_O} \sum_{k=1}^{K} \omega_k \|\tilde{R}_k - R\|^2 \quad (12)$$

In view of (8), this means finding:

$$\arg\min_{R \in C_H \cap S_O} \|\bar{R} - R\|^2$$

Appendix 2-b proves that, if $C$ is a cone and $S$ the unit sphere of a euclidean space:

$$\arg\min_{y \in C \cap S} \|y - x\|^2 = \frac{\hat{y}}{\|\hat{y}\|}, \text{ where } \hat{y} = \arg\min_{y \in C} \|y - x\|^2$$

As a consequence, the solution of (12) is:

$$\frac{\hat{R}_H}{\|\hat{R}_H\|} \text{ with: } \hat{R}_H = \arg\min_{R \in C_H} \|\bar{R} - R\|^2 \quad (13)$$

It is a classical matrix-algebra result that $\hat{R}_H = U \hat{\Lambda}_H U'$, where $\hat{\Lambda}_H$ is the diagonal matrix the diagonal elements of which are the largest $H$ eigenvalues of $\bar{R}$ sorted by descending order, all other elements being zero, and $U$ is the orthogonal matrix the column-vectors of which are the eigenvectors of $\bar{R}$ sorted by decreasing eigenvalue.

With $H=rank(\bar{R})$, one thus gets back the previous $S_O$-constrained average $\frac{\bar{R}}{[\![\bar{R}]\!]}$.

Interpreting $\hat{R}_H$:

When the resultants code data-arrays each endowed with a specific metric, i.e. $\forall k, \tilde{R}_k = r(X_k, M_k)$, then:

$$\bar{R} = \sum_k R_{X_k, N_k} \text{ with } N_k = \frac{\omega_k}{[\![R_{X_k, M_k}]\!]} M_k$$

Thus, its first $H$ eigenvectors are those of the PCA of array $X=[X_1,...,X_K]$ with block-diagonal metric $N=diag(N_1,...,N_K)$ and weights $W$.

$\hat{R}_H$ thus appears to be a structural trade-off between the data-structures encoded in the resultants, e.g. the $X_k$'s. This trade-off contains information increasing with $H$, and eventually, when $H=rank(\mathbf{R})$, captures the whole information of $X$. To better see how $\hat{R}_H$ works, let us first assume that $H=rank(\mathbf{R})$. Then, $\hat{R}_H \propto R_{X,N} = XNX'W$. Now consider a subspace $E \subset \mathbb{R}^n$. We have:

$$\cos(\hat{R}_H, \Pi_E) = \frac{[R_{X,N} | \Pi_E]}{[\![R_{X,N}]\!] [\![\Pi_E]\!]}$$

Now:

$$[R_{X,N} | \Pi_E] = tr(XNX'W\Pi_E) = tr(XNX'W\Pi_E^2) = tr(\Pi_E XNX'W\Pi_E)$$
$$= tr(\hat{X}_E N \hat{X}_E' W) \text{ with } \hat{X}_E := \Pi_E X$$

Besides: $[\![R_{X,N}]\!] = \sqrt{tr(R_{X,N}^2)} = \sqrt{\sum_j \lambda_j^2}$, where the $\lambda_j$'s are the eigenvalues of $R_{X,N}$, and $[\![\Pi_E]\!] = \sqrt{(dim\, E)}$. So, up to a multiplicative constant, $\cos(\hat{R}_H, \Pi_E)$ measures the part of the inertia of $X$ that is along $E$. Note the presence of $dim\, E$ in the denominator of this constant has the effect of penalising the dimensionality of the subspace $E$ used to capture the inertia, which is easily defendable.

With $H < rank(\bar{R})$, $\hat{R}_H$ only retains the inertia of $(X, N)$ captured by its first $H$ principal components, but the heuristic interpretation is just the same: if the $rank(\bar{R}) - H$ components dropped can be considered plain noise, then, up to a multiplicative constant, $\cos(\hat{R}_H, \Pi_E)$ measures the part of the structural information of $X$ that is along $E$.

From a practical standpoint, it is important to make a "good" choice of $H$. Several criteria may be considered:

1. The most straightforward criterion is $c_1 = \frac{trace(\hat{R}_H)}{trace(\bar{R})}$, i.e. the pct of information of the weighted resultants captured by $\hat{R}_H$.

2. $c_2$ = Cattell's criterion, based on the second-order differences of eigenvalues.

    Indeed, $c_1$ and $c_2$ are naive in that they do not take into account the sample fluctuations of the ranked eigenvalues.

3. [Saporta et al. 1993], and [Saporta et al. 1999] have proposed interesting criteria for identifying the number of significant components in a PCA or a MCA, taking into account the fluctuations of the ranked eigenvalues. These criteria can be used when clustering variables of the same nature, quantitative or categorical (cf. section 4), as in [Derquenne 2016].

4. A more empirical and more robust idea would be to use bootstrapping so as to

## 3.2. Averaging resultants using the geodesic distance  $\delta$

### 3.2.1. The geodesic rank-H average

Let $R_H$ be a rank- $H$ normed resultant. We have:
$$R_H = U \Lambda U'W \quad , \text{where}$$
$$U = [u_1, \ldots, u_H], U'WU = I_H, \Lambda = diag(\lambda) \quad \text{with} \quad \lambda = (\lambda_h; 1 \leq h \leq H) \quad , \text{and}$$
$$\|R_H\|^2 = 1 \Leftrightarrow tr(\Lambda^2) = 1 \Leftrightarrow \|\lambda\|^2 = 1 \quad .$$

In view of (7), the rank- $H$ geodesic average may be defined as:
$$\arg\min_{R \in C_H \cap S_o} \sum_{k=1}^K \omega_k \delta^2(\tilde{R}_k, R) = \arg\min_{R \in C_H \cap S_o} \sum_{k=1}^K \omega_k (\cos^{-1}([\tilde{R}_k|R]))^2$$
$$= \arg\max_{R \in C_H \cap S_o} -\sum_{k=1}^K \omega_k (\cos^{-1}([\tilde{R}_k|R]))^2 \qquad (14),$$
$$\text{with} \quad [\tilde{R}_k|R] = tr(\tilde{R}_k U \Lambda U'W) = tr(U'W\tilde{R}_k U \Lambda)$$

### 3.2.2. A general program and characterization of critical points

Let $g$ be a real function of $\lambda$ and $U$. We shall write:
$$\gamma(\lambda, U) = \frac{\partial g(\lambda, U)}{\partial \lambda} \in \mathbb{R}^H \quad ; \quad \Gamma(\lambda, U) = \frac{\partial g(\lambda, U)}{\partial U} \in \mathbb{R}^{n \times H}$$

We now tackle the program:
$$\max_{\substack{\lambda \in \mathbb{R}^H, \|\lambda\|^2 = 1 \\ U \in \mathbb{R}^{n \times H}, U'WU = I_H}} g(\lambda, U)$$

Appendix 2-c shows that a point that satisfies:
$$\lambda = \frac{\gamma(\lambda, U)}{\|\gamma(\lambda, U)\|} \quad ; \quad U = W^{-1} \Gamma(\lambda, U)(\Gamma(\lambda, U)'W^{-1}\Gamma(\lambda, U))^{-\frac{1}{2}} \quad (15)$$
is a critical point of this program.

### 3.2.3. An ascending iteration to a critical point

Appendix 2-d shows that:
$$\forall G \in \mathbb{R}^{n \times H} : V = W^{-1} G (G'W^{-1}G)^{-\frac{1}{2}} = \arg\max_{U'WU = I_H} tr(U'G) \quad (16)$$

Using (16), we build up the following iteration and show that it follows a direction of ascent for $g$:
$$\lambda^{[t]} = \frac{\gamma^{[t-1]}}{\|\gamma^{[t-1]}\|} \quad , \quad U^{[t]} = W^{-1} \Gamma^{[t-1]}(\Gamma^{[t-1]}{}'W^{-1}\Gamma^{[t-1]})^{-\frac{1}{2}} \quad (17)$$
$$\text{where} \quad \gamma^{[t-1]} = \gamma(\lambda^{[t-1]}, U^{[t-1]}) \quad \text{and} \quad \Gamma^{[t-1]} = \Gamma(\lambda^{[t-1]}, U^{[t-1]})$$

If this iteration reaches a fixed point, it is obvious that it satisfies (15).
Besides, let us show that it follows a direction of ascent. Firstly:
$$\langle \lambda^{[t]} - \lambda^{[t-1]} | \gamma^{[t-1]} \rangle = \langle \lambda^{[t]} - \lambda^{[t-1]} | \lambda^{[t]} \rangle \|\gamma^{[t-1]}\|$$
Now, $\langle \lambda^{[t]} - \lambda^{[t-1]} | \lambda^{[t]} \rangle = \|\lambda^{[t]}\|^2 - \langle \lambda^{[t-1]} | \lambda^{[t]} \rangle = 1 - \cos(\lambda^{[t-1]}, \lambda^{[t]}) \geq 0$ .

Secondly:

$$tr((U^{[t]}-U^{[t-1]})'\Gamma^{[t-1]})=tr((W^{-1}\Gamma^{[t-1]}(\Gamma^{[t-1]}{}'W^{-1}\Gamma^{[t-1]})^{-\frac{1}{2}}-U^{[t-1]})'\Gamma^{[t-1]})\geq 0 \;,$$

in view of (16).

Of course, picking a point on a direction of ascent does not guarantee that $g$ actually increases, since we may "go too far" in this direction. But if we stay "close enough" to the current starting point on the arc going from $(\lambda^{[t-1]}, U^{[t-1]})$ to $(\lambda^{[t]}, U^{[t]})$, we can guarantee that $g$ increases. In practice, we suggest to look for the maximum of $g$ on the arc, i.e. to solve the following scalar program:

$$\max_{\tau\in[0,1]} g(R_\tau^{[t-1]}) \;,\text{ where: } R_\tau^{[t-1]}=\frac{R^{[t-1]}+\tau(R^{[t]}-R^{[t-1]})}{\|R^{[t-1]}+\tau(R^{[t]}-R^{[t-1]})\|},\; R^{[t]}=U^{[t]}\Lambda^{[t]}U^{[t]}{}'W$$

### 3.2.4. The program for the rank-H geodesic average

In view of (14), the rank-H geodesic average is found applying the aforementioned iteration to function:

$$g(\lambda, U)=-\sum_{k=1}^{K}\omega_k(\cos^{-1}(tr(U'W\tilde{R}_k U \Lambda)))^2$$

*Derivatives:*

Let $\forall k: A_k=W\tilde{R}_k$ symmetric and: $h_k(U,\lambda)=tr(U'A_k U \Lambda)$, with $\Lambda=diag(\lambda)$.

Differentiating $h_k$ with respect to $\lambda$, we get:

$$dh_k(U,\lambda)=tr(U'A_k U d\Lambda)=\langle\eta_k|d\lambda\rangle \text{ with } \eta_k=(u_h'A_k u_h; 1\leq h\leq H)$$

Hence: $\dfrac{\partial h_k(U,\lambda)}{\partial \lambda}=\eta_k$.

In the same manner, differentiating $h_k$ with respect to $U$, we get:

$$dh_k(U,\lambda)=tr(dU'A_k U \Lambda)+tr(U'A_k dU \Lambda)=tr((2A_k U \Lambda)' dU)$$

Hence: $\dfrac{\partial h_k(U,\lambda)}{\partial U}=2 A_k U \Lambda$.

This straightforwardly yields:

$$\gamma(U,\lambda)=\frac{\partial g(U,\lambda)}{\partial \lambda}=\sum_{k=1}^{K}\omega_k\frac{1}{(1-h_k^2(U,\Lambda))^{\frac{1}{2}}}\eta_k$$

$$\Gamma(U,\lambda)=\frac{\partial g(U,\lambda)}{\partial U}=\left[2\sum_{k=1}^{K}\omega_k\frac{1}{(1-h_k^2(U,\Lambda))^{\frac{1}{2}}}A_k\right]U\Lambda$$

*Initialization:*

Since distances $d$ and $\delta$ are locally equivalent, the iteration for finding the rank-$H$ geodesic average should be initialized with the plain rank-H average.

# 4. Extending the K-means-type variable-set-clustering methods

## 4.1. Extension using the euclidean distance d

In view of section 3.1., the methods mentioned in section 1.4 which cluster variables around their first PC appear to be no other than the K-means method applied to the normed resultants coding the variables on $S_O$, taking the rank-1 $S_O$-constrained average of each cluster as centroid. Indeed, let $\forall k: \tilde{R}_k = r(X_k, M_k)$, where $X_k$ codes the $k^{th}$ variable and $M_k = (X_k'WX_k)^{-1}$, so that $\tilde{R}_k$ is the normed projector on $\langle X_k \rangle$. Besides, let $f_l f_l' W$ be the normed rank-1 average of variable-cluster $c_l$ ($f_l$ is then the normed first PC of $c_l$). When associating each resultant with the closest centroid, the K-means looks for:

$$\arg\min_{1 \leq l \leq L} \| \tilde{R}_k - f_l f_l' W \|^2 = \arg\max_{1 \leq l \leq L} [\tilde{R}_k | f_l f_l' W]$$

Now:

$$[\tilde{R}_k | f_l f_l' W] = tr(\tilde{R}_k f_l f_l' W) ,$$
$$\text{and} \quad tr(\tilde{R}_k f_l f_l' W) = f_l' W \tilde{R}_k f_l$$
$$= f_l' W \Pi_{\langle X_k \rangle} f_l = \cos^2(f_l, \langle X_k \rangle) ,$$

which we have seen in section 2.2. was the classical bivariate square correlation between $f_l$ and $X_k$ appropriate to the type of the latter.

It should be noted that coefficient $\varpi_k = \dfrac{1}{\|X_k M_k X_k' W\|} = 1$ when $X_k$ is a numerical variable, and $\varpi_k = \dfrac{1}{\sqrt{m_k - 1}}$ when $X_k$ is categorical with $m_k$ levels. Thus, categorical variables are re-weighted according to the dimension of their subspace. This can be and was defended by [Qannari et al. 1998].

The resultant-encoding and the previous results on the normed rank-$H$ average allow us to unify and extend these methods as follows.

Firstly we need not consider that the $X_k$'s code isolated variables, but variable-*structures*, i.e. variable-subsets endowed each with a specific metric $M_k$ specifying the way we look at the structure of $X_k$. Secondly, we may consider using a rank-$H$ average instead of the mere rank-1 one. Rank $H$ can evolve throughout the algorithm, according to the part of the cluster's structure its average should represent. Indeed, a narrow cluster of highly-correlated $X_k$ structures may be correctly represented by its 1$^{st}$ PC, but a more scattered one would need more PC's. $H$ allows to take into account the dispersion of each cluster. This is similar to what is being done when clustering statistical units and the within-cluster variance matrix is re-estimated on every step, and used when re-allocating units to clusters.

Applying the K-means procedure, we then get the following algorithm:

Step 0: Initialise the $L$ clusters at random
Current step $t$ :
 1. Calculate the average $R_l$ of the resultants in each cluster $c_l$, diagonalise it and choose $H_l$ appropriately. Calculate the $H_l$-average $\hat{R}_{l,H_l} = U_l \hat{\Lambda}_{l,H_l} U_l'$ and norm it, to get the centroid of $c_l$ : $\tilde{R}_{l,H_l}$.

2. Update clusters by re-affecting each $\widetilde{R}_k$ to $c_{l(k)}$ where:
$$l(k) = \arg \min_{1 \leq l \leq L} [\tilde{\bar{R}}_{l,H_l} | \tilde{R}_k] .$$
3. If clusters are unchanged, stop. Else, resume 1.

It can be seen here that $H_l$ should be recalculated each time cluster $c_l$ changes. So it would be better to use a simple criterion, like $c_1$ or $c_2$.

### 4.2. Extension using the geodesic distance $\delta$

The K-means algorithm remains unchanged while using $\delta$ instead of $d$.

### 4.3. Summarising the clusters

A cluster $c_l$ can be summarized by its rank-$H$ average according to either $d$ or $\delta$. This average may be analysed via diagonalisation, which, in the case of $d$, amounts to performing usual PCA on $X$.

As to the cluster's dispersion about its average, it may for instance be summarised by measuring the closeness of each variable $x^k \in c_l$ having resultant $R_k$, to the cluster's average $\bar{R}_{l,H_l}$, and producing the box-blot of these measures. This closeness may for instance be measured through $\cos(R_k | \bar{R}_{l,H_l}) = [\tilde{R}_k | \tilde{\bar{R}}_{l,H_l}]$. Recall that when $x^k$ is quantitative, this is equal to the part of information in $\bar{R}_{l,H_l}$ captured by $x^k$, and that when $x^k$ is categorical with $r$ levels, this cosine is equal to the part of information in $\bar{R}_{l,H_l}$ captured by $x^k$ divided by $\sqrt{r-1}$.

One may alternatively focus on the distance $d$ or $\delta$ of variables to the cluster's average.

## 5. Numerical results

### 5.1. Testing the method on simulated data

#### 5.1.1. Simulation scheme for one variable-sample

We simulate through the following process three variable clusters $A$, $B$ and $C$ of respectively 9, 6 and 6 variables, such that $A$ has, up to some noise, an isotropic distribution in a plane of $\mathbb{R}^n$, and $B$, $C$ are bundles distributed each about a line:

1. Simulate as follows four latent variables: $\{\xi_1, \xi_2\}$ to span $A$'s plane, and $\xi_3$, $\xi_4$ to be the central directions of $B$ and $C$ respectively (cf. Fig. 1):
    a) Simulate 4 independent $N(0,1)$ variables in $\mathbb{R}^n$ : $\xi_1, \xi_2, \xi_3, \xi_4$ and centre them.
    b) Orthogonalise $\xi_2$ with respect to $\xi_1$ by updating it as follows:
    $$\xi_2 = \xi_2 - \frac{\xi_2' \xi_1}{\xi_1' \xi_1} \xi_1 .$$
    c) Standardise all these variables.
    d) To tune the correlation of cluster $B$ to cluster $A$, we put:
    $\xi_3 = \xi_2 \cos\beta + \xi_3 \sin\beta$ , with $\beta = \frac{\pi}{2}$ , $\beta = \frac{\pi}{3}$ , and $\beta = \frac{\pi}{4}$ . When $\beta = \frac{\pi}{2}$ , A and B are uncorrelated, and the farther $\beta$ from $\frac{\pi}{2}$ , the more "structural confusion" between them.
    e) Standardise $\xi_3$ .

2. Simulate as follows the actual numerical variables in the clusters:
   a) Simulate $p$ independent normal noise variables $\varepsilon^j \sim N(0, \sigma^2)$, where $\sigma^2$ is the amplitude of the noise and tunes the width of the bundle about its central latent subspace.
   b) Cluster $A$ : sample 7 independent values of $\alpha$ from the uniform distribution on $[0; 2\pi]$ and, for each, set: $x^j = \cos\alpha_j \xi_1 + \sin\alpha_j \xi_2 + \varepsilon^j$
   c) Cluster $B$ : Set $x^j = \xi_3 + \varepsilon^j$, $j \in \{8, \ldots, 12\}$ .
   d) Cluster $C$ : Set $x^j = \xi_4 + \varepsilon^j$, $j \in \{13, \ldots, 17\}$ .
3. Simulate as follows categorical variables in the clusters: $x^{18}, x^{19}, x^{20}, x^{21}$ are categorical variables derived from $\xi_1, \xi_2, \xi_3, \xi_4$ through quantile partition:
   a) Let $q_{1,\alpha}, q_{2,\alpha}, q_{2,\alpha}, q_{4,\alpha}$ denote the $\alpha$-order quantiles respectively issued from $\xi_1, \xi_2, \xi_3, \xi_4$ .
   b) The 5-label variable $x^{18}$ issued from $\xi^1$ is given by:
      $x_i^{18} = j, j \in \{1, \ldots, 5\} \Leftrightarrow \xi_i^1 \in ]q_{1, \frac{j-1}{5}}, q_{1, \frac{j}{5}}]$
   c) The 5-label variables $x^{19}, x^{20}, x^{21}$ issued from $\xi^2, \xi^3, \xi^4$ are obtained likewise.
   d) Variables $x^{18}$ and $x^{19}$ are assigned to cluster $A$ , variable $x^{20}$ to cluster $B$ and variable $x^{21}$ to cluster $C$ .

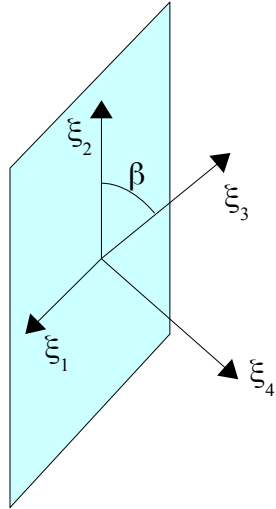

*Figure 1: the four latent variables underlying the clusters*

### 5.1.2. Results

The simulation scheme was used 100 times for each choice of $(n, \beta, \sigma^2)$ , and our K-means algorithm was used with $K=3$ being the true number of clusters, known here, and 5 values of rank $H$ corresponding to the following values $\theta$ of trace-ratio $c_1$ (cf section 3.1.3.): $\theta \in \{0, .25, .5, .75, 1\}$ . When $\theta = 0$ , $H = 1$ , and when $\theta = 1$ , $H$ is the rank of the cluster's average resultant. Then, each of the yielded partitions was compared to the true partition through the Rand index (number of variable-pairs belonging to the same cluster in one partition but not the other, divided by the number of variable-pairs belonging to the same cluster in at least one partition). The lower the Rand index, the more adequate the clustering, in our case. The rand indexes were averaged across the 100 samples and a standard deviation was calculated. Results are given in table 1. They show several things:

1. A smaller $\sigma^2$, all other parameter values kept equal, brings better results, unsurprisingly.
2. Starting from $\frac{\pi}{2}$, decreasing $\beta$ increases the structural confusion between clusters $A$ and $B$, which raises the Rand index.
3. Increasing $n$ improves the accuracy of the clustering, by lowering the probability that the sample-covariance of two independently generated variables exceed a given threshold.
4. The really interesting result is that increasing $\theta$ increases the accuracy of the clustering. The improvement is all the more notable as the noise and confusion parameters are low. E.g. for $n=40, \beta=\frac{\pi}{2}, \sigma^2=0.1$, increasing $\theta$ from 0 to 1 decreases the Rand index from 7.5% to 0.4%. Indeed, when noise and confusion are close to zero, the cluster's average resultant captures the true structural dimensionality of the cluster, and it is of essence to use this full dimensionality as that of the $H$-average. In our example, it is clear that cluster $B$ being almost uniformly distributed on a plane, it could only be clustered about a single dimension at great loss.

|  | $\beta = \pi/4$ | $\beta = \pi/3$ | $\beta = \pi/2$ |
|---|---|---|---|
| n=30 $\sigma^2 = 0.1$ | | | |
| $\theta = 0.$ | 0.188 (0.058) | 0.151 (0.065) | 0.088 (0.059) |
| $\theta = 0.25$ | 0.186 (0.059) | 0.148 (0.067) | 0.087 (0.054) |
| $\theta = 0.5$ | 0.164 (0.054) | 0.065 (0.068) | 0.023 (0.048) |
| $\theta = 0.75$ | 0.161 (0.062) | 0.059 (0.058) | 0.021 (0.044) |
| $\theta = 1.$ | 0.134 (0.055) | 0.046 (0.052) | 0.008 (0.036) |
| n=30 $\sigma^2 = 0.15$ | | | |
| $\theta = 0.$ | 0.184 (0.052) | 0.141 (0.068) | 0.101 (0.054) |
| $\theta = 0.25$ | 0.182 (0.051) | 0.140 (0.071) | 0.100 (0.057) |
| $\theta = 0.5$ | 0.165 (0.049) | 0.110 (0.069) | 0.058 (0.064) |
| $\theta = 0.75$ | 0.163 (0.051) | 0.106 (0.072) | 0.053 (0.061) |
| $\theta = 1.$ | 0.149 (0.065) | 0.091 (0.078) | 0.036 (0.055) |
| n=40 $\sigma^2 = 0.1$ | | | |
| $\theta = 0.$ | 0.187 (0.053) | 0.131 (0.062) | 0.075 (0.056) |
| $\theta = 0.25$ | 0.186 (0.060) | 0.130 (0.063) | 0.074 (0.051) |
| $\theta = 0.5$ | 0.164 (0.057) | 0.782 (0.069) | 0.012 (0.057) |
| $\theta = 0.75$ | 0.159 (0.061) | 0.755 (0.049) | 0.010 (0.050) |
| $\theta = 1.$ | 0.124 (0.059) | 0.041 (0.058) | 0.004 (0.039) |
| n=40 $\sigma^2 = 0.15$ | | | |
| $\theta = 0.$ | 0.187 (0.055) | 0.154 (0.063) | 0.091 (0.052) |
| $\theta = 0.25$ | 0.186 (0.053) | 0.152 (0.067) | 0.089 (0.058) |
| $\theta = 0.5$ | 0.176 (0.052) | 0.136 (0.058) | 0.085 (0.059) |
| $\theta = 0.75$ | 0.167 (0.057) | 0.135 (0.071) | 0.079 (0.060) |
| $\theta = 1.$ | 0.160 (0.062) | 0.116 (0.066) | 0.052 (0.057) |

*Table 1: Numerical results of clustering on simulated data: mean Rand index according to $n, \beta, \sigma^2, \theta$ (standard deviation in parentheses)*

## 5.2. Application to real data

We apply our method to the wine data of the *PCAmixdata* R-library. The data contains 31 variables: 29 quantitative variables (sensory scores, as astringency, bouquet, colour...) and 2 categorical variables (soil-type and label). These variables describe $n = 21$ wines.

We first calculate the normed resultant operator $\tilde{R}_k$ associated with each variable $x^k$. Then, we calculate the normed rank-$H$ average of these operators: $\tilde{R}_H$, for $H$ going from 1 to 21, and for each value of $H$, calculate the "geodesic inertia" of variables about the average:

$$D_H = \sum_k \delta^2(\tilde{R}_k, \tilde{R}_H)$$

The results are given in table 2. From $H=2$ on, we see that the performance of $\tilde{R}_H$ remains almost constant. This is consistent with what a mixed data PCA reveals: the plane spanned by the first 2 components contains 65% of the inertia and the scree-plot shows a drop between components 2 and 3, with a slow regular decrease of the eigenvalues from 3 to 21.

| $H$ | 1 | 2 | 3 | 4 | ... | 21 |
|---|---|---|---|---|---|---|
| $D_H$ | 36.00564 | 32.55528 | 32.06235 | 31.82901 | ... | 31.42671 |

*Table 2: Geodesic inertia of variables about their rank-$H$ geodesic average, according to $H$.*

The geodesic K-means procedure was launched for values of $K$ ranging from 2 to 5, with a value of $H \in \{1,2\}$ being the smallest such that the rank-$H$ average (centroid) of the cluster would capture at least 50% of its information. The ratio *Between-inertia/Total inertia* was computed for each produced partition (cf fig 1).

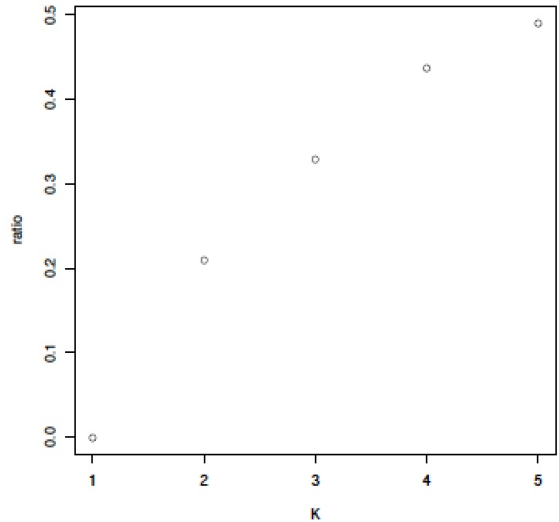

*Figure 1: Between-inertia/Total inertia ratio for K = 1 to 5.*

According to second-order differences between consecutive values, partitions in respectively 2 and 4 clusters should be preferred.

*Partition into 2 clusters:*

The cosine between the two cluster-centroids is 0.376, which means that the clusters are well-separated. The contents of the clusters and cosine of each variable with the centroid of its cluster are given in table 3. The variables with the highest cosine are those which best illustrate their cluster (in bold-type).

| Cluster 1 | | Cluster 2 | |
|---|---|---|---|
| Variable $k$ | $\cos(R_k|\bar{R}_{1,H_1})$ | Variable $k$ | $\cos(R_k|\bar{R}_{2,H_2})$ |
| AromaBS | 0.529 | OdorBS | 0.659 |
| FruityBS | 0.669 | FlowerBS | 0.646 |
| **VisualINT** | **0.744** | **SpiceBS** | **0.831** |
| Nuance | 0.647 | **OdorINT** | **0.747** |
| SurfFEEL | 0.506 | Flower | 0.737 |
| **OdorQUA** | **0.763** | Spice | 0.154 |
| **Fruity** | **0.748** | **Phenolic** | **0.808** |
| Plante | 0.558 | Astringency | 0.705 |
| **AromaINT** | **0.862** | Bitterness | 0.379 |
| **AromaPER** | **0.832** | Label | 0.405 |
| **AromaQUA** | **0.903** | Soil | 0.52 |
| **AttackINT** | **0.770** | | |
| **Acidity** | **0.779** | | |
| Alcohol | 0.371 | | |
| Balance | 0.517 | | |
| Smooth | 0.704 | | |
| Intensity | 0.525 | | |
| Harmony | 0.391 | | |
| OverallQUA | 0.599 | | |
| Typical | 0.562 | | |

Table 3: *Contents of clusters and cosine of each variable with the centroid of its cluster for the 2-cluster partition.*

**Partition into 4 clusters:**

Clusters 1 and 2 were found to be best represented by rank-2 centroids, whereas all other clusters were by a rank-1 centroid, i.e. a single dimension.

The cosines between the cluster-centroids are given in table 4. The clusters are well-separated, the best separated being clusters 2 and 4, and the least being clusters 3 and 4.

| Cosine | Cluster 1 | Cluster 2 | Cluster 3 | Cluster 4 |
|---|---|---|---|---|
| Cluster 1 | 1 | 0.2586 | 0.1965 | 0.2829 |
| Cluster 2 | | 1 | 0.3970 | 0.0701 |
| Cluster 3 | | | 1 | 0.6170 |
| Cluster 4 | | | | 1 |

Table 4: *Cosines between cluster-centroids for the 4-cluster partition.*

Multidimensional scaling (MDS) was performed from the distance matrix between the clusters' centroids. The first principle plane is shown on figure 2.

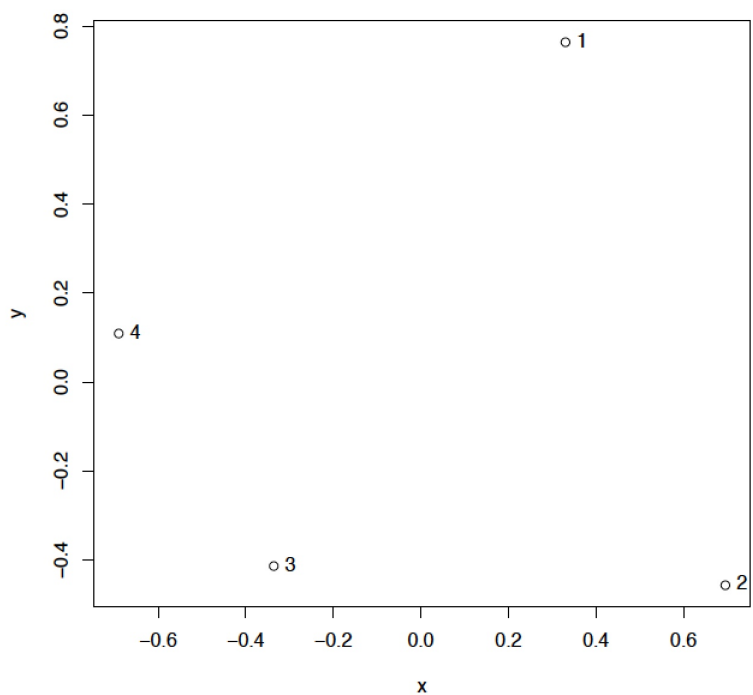

*Figure 2: Principal MDS plane of cluster-centroids.*

The contents of the clusters and cosine of each variable with the centroid of its cluster are given in table 5.

| Cluster 1 | | Cluster 2 | | Cluster 3 | | Cluster 4 | |
|---|---|---|---|---|---|---|---|
| Variable | Cos with centroid | Variable | Cos with centroid | Variable | Cos with centroid | Variable | Cos with centroid |
| FlowerBS | 0.557 | OdorBS | 0.572 | VisualINT | 0.628 | AromaBS | 0.583 |
| SpiceBS | 0.448 | OdorINT | 0.547 | **Nuance** | **0.777** | FruityBS | 0.627 |
| **Flower** | **0.860** | **Phenolic** | **0.761** | **SurfFEEL** | **0.826** | OdorQUA | 0.593 |
| **Spice** | **0.842** | **Bitterness** | **0.756** | **AromaINT** | **0.862** | Fruity | 0.612 |
| **Acidity** | **0.892** | **Label** | **0.838** | **AromaPER** | **0.793** | **Plante** | **0.898** |
| **Soil** | **0.830** | | | AttackINT | 0.677 | AromaQUA | **0.767** |
| | | | | **Astringency** | **0.814** | Balance | 0.595 |
| | | | | Alcohol | 0.639 | **Smooth** | **0.802** |
| | | | | Intensity | 0.651 | **OverallQUA** | **0.886** |
| | | | | Harmony | 0.868 | Typical | 0.869 |

*Table 5: Contents of clusters and cosine of each variable with the centroid of its cluster for the 4-cluster partition.*

## 6. Conclusion

In this work, we have provided a unified geometric framework for variable-clustering and extended it to variable-subsets. We have proposed to represent every subset of variables, by a normed resultant operator on the unit sphere of $\mathbb{R}^{n \times n}$. Two distances can be used on this sphere: the chord distance and the geodesic one. Using either of them does not lead to great differences, and the chord-distance is computationally much cheaper. We have proposed the notion of rank-$H$ average of resultants, and showed that considering a high $H$ may be of the essence when calculating a cluster's average, for fear that too much of the cluster's valuable dimensional information be lost.

# References

Robert, P.; Escoufier, Y. (1976). *A Unifying Tool for Linear Multivariate Statistical Methods: The RV-Coefficient*. Applied Statistics 25 (3): 257–265.

Bry X. (2001) : *Une autre approche de l'Analyse Factorielle : l'Analyse en Résultantes Covariantes* – RSA vol. XLIX (3), pp. 5-38.

Chavent, M., Kuentz-Simonet, V., Liquet, B., and Saracco, J. (2012). *Clustofvar : An R package for the clustering of variables*. Journal of Statistical Software, 50(13) :1–16.

Escoufier Y. (1970), *Échantillonnage dans une population de variables aléatoires réelles*, Publications de l'Institut de Statistique de l'Université de Paris, Vol XIX, fasc. 4, p 1-47.

Gomes P. (1987), *Distribution de Bingham sur la n-sphere: une nouvelle approche de l'analyse factorielle*. PhD Thesis, Université Montpellier 2.

Gomes A. (1993), *Reconnaissance de mélanges de lois de Bingham: application à la classification de variables*. PhD Thesis, Université Montpellier 2.

Qannari, E.M., Vigneau E., Courcoux Ph., (1998): *Une nouvelle distance entre variables. Application en classification*. RSA, XLVI, 2 .

Vigneau E., Qannari E.M., Sahmer K., Ladiray D., (2006): *Classification de variables autour de composantes latentes*, RSA, LIV, 1, pp. 27-45

Vigneau E., Qannari E.M. (2003), *Clustering of variables around latent components*, Communications in Statistics : Simulation and Computation, 32 (4), 1131-1150.

Abdallah H., Saporta G. (1998), *Classification d'un ensemble de variables qualitatives*, Revue de Statistique Appliquée, XLVI(4), 5-26.

Saracco J., Chavent M., Kuentz V. (2010), *Clustering of categorical variables around latent variables*, Cahiers du GREThA UMR CNRS 5113, février 2010, Université Bordeaux 4.

Chavent M., Kuentz V., Liquet B., Saracco J. (2012) *ClustOfVar: An R Package for the Clustering of Variables*, Journal of Statistical Software, Vol. 50, 13.

Chavent M., Kuentz V., Saracco J. (2010) *A Partitioning Method for the Clustering of Categorical Variables*, Proceedings of the 11th IFCS Biennial Conference and 33rd Annual Conference of the Gesellschaft für Klassifikation e.V., Dresden, march 2009.

Vigneau E., Qannari E.M., Sahmer K., Ladiray D. (2006) *Classification de variables autour de composantes latentes*, Rev. Statistique Appliquée, 2006, LIV (1), 27-45

Dortet-Bernadet J.-L., Wicker N., (2008) *Model-based clustering on the unit sphere with an illustration using gene expression profiles*, Biostatistics (2008), 9, 1, pp. 66–80.

Banerjee A., Dhillon I. S., Ghosh J., Sra S. (2005), *Clustering on the Unit Hypersphere using von Mises-Fisher Distributions*, Journal of Machine Learning Research 6 (2005) 1345–1382

Palla K., Knowles D. A., Ghahramani Z. (2012), *A nonparametric variable clustering model*, NIPS 2012 Conference.

G.Saporta, N.Tambrea (1993): *About the selection of the number of components in correspondence analysis*, in J.Janssen et C.H.Skiadas, eds. Applied Stochastic Models and Data Analysis, World Scientific, p. 846-856.

G. Saporta (1999): *Some simple rules for interpreting outputs of principal components and correspondence analysis*. In ASMDA99, IX International Symposium on Applied Stochastic Models and Data Analysis, Lisbon, Portugal, 14-17 June 1999.

Derquenne (2016): *Classification de variables : Une approche à double critères contrôlés dynamiques*, 48èmes Journées de Statistique, Montpellier, May 2016.

# Appendix 1: Notations

The space spanned by a set of vector $X$ will be denoted $\langle X \rangle$.

The unit sphere of $\mathbb{R}^n$ will be denoted $S_{n-1}$.

The orthogonal projector (with respect to a given metric) on space $E$ will be denoted $\Pi_E$. If $E=\langle X \rangle$, notation $\Pi_{\langle X \rangle}$ will be simplified to $\Pi_X$.

$\forall X$ (respectively $Y$) an $n \times p$ (resp. $n \times q$) matrix, $[X,Y]$ denotes the $n \times (p+q)$ matrix obtained by juxtaposing the columns of $X$ and $Y$ in that order.

# Appendix 2: Proofs

**a)** $\Phi^2(X,Y) = tr(W^{-1} \Pi_X' W \Pi_Y)$ :

Let $X$ and $Y$ be two categorical variables with $q$ and $r$ levels respectively, and let **X** and **Y** denote their respective matrices of $q$ (resp. $r$) uncentred indicator-variables, and $X$ and $Y$ their respective matrices of $q-1$ (resp. $r-1$) centred indicator-variables. We have:

$$\langle X \rangle = \langle \mathbf{X} \rangle \cap \langle \mathbf{1} \rangle^\perp \; ; \; \langle Y \rangle = \langle \mathbf{Y} \rangle \cap \langle \mathbf{1} \rangle^\perp$$

And so:

$$\Pi_X = \Pi_\mathbf{X} - \Pi_\mathbf{1} \; ; \; \Pi_Y = \Pi_\mathbf{Y} - \Pi_\mathbf{1} \quad \text{(a)}$$

Besides, it can easily be shown that:

$$tr(W^{-1} \Pi_\mathbf{X}' P \Pi_\mathbf{Y}) = tr(MM') \text{ (b), where}$$
$$M = (\mathbf{X}' W \mathbf{X})^{-1/2} \mathbf{X}' W \mathbf{Y} (\mathbf{Y}' W \mathbf{Y})^{-1/2}$$

We have:

$$M = (m_{ij})_{i,j} \quad \text{with} \quad m_{ij} = \frac{\pi_{ij} - \pi_{i.} \pi_{.j}}{\sqrt{\pi_{i.} \pi_{.j}}} \quad \text{(c)}$$

From (a), (b) and (c), we get:

$$tr(W^{-1} \Pi_X' W \Pi_Y) = \sum_{\substack{i=1 \text{ to } q \\ j=1 \text{ to } r}} \frac{(\pi_{ij} - \pi_{i.} \pi_{.j})^2}{\pi_{i.} \pi_{.i}} = \Phi^2(X,Y)$$

**b)** $\arg\min_{y \in C \cap S} \|x-y\|^2 = \frac{\hat{y}}{\|\hat{y}\|}$, where $\hat{y} = \arg\min_{y \in C} \|x-y\|^2$

*Proof*:

Let $\forall y, y^0 = \frac{y}{\|y\|}$.

$$\forall x, \|x - y^0\|^2 = \|x\|^2 + \|y^0\|^2 - 2\langle x | y^0 \rangle$$

So:

$$\arg\min_{y \in C \cap S} \|x-y\|^2 = \arg\max_{y \in C \cap S} \langle x | y \rangle$$
$$= \arg\max_{y \in C \cap S} (\|x\|^2 \cos^2(x,y)) = \arg\min_{y \in C \cap S} (\|x\|^2 \sin^2(x,y))$$

$$= \hat{y}^0, \text{ forall } \hat{y} = \arg\min_{y \in C} (\|x\|^2 \sin^2(x, y)) \quad (C \text{ being a cone}, y^0 \in C)$$

For instance:
$$\hat{y} = \arg\min_{y \in C} \|x - y\|^2 = \Pi_C(x)$$

**c)** Characterizing critical points of the program $\max_{\substack{\lambda \in \mathbb{R}^H, \|\lambda\|^2 = 1 \\ U \in \mathbb{R}^{n \times H}, U'WU = I_H}} g(\lambda, U)$

The corresponding lagrangian is:
$$L = g(\lambda, U) - \frac{\mu}{2}(\lambda'\lambda - 1) - \frac{1}{2} tr((U'WU - I_H) M),$$

where $\mu \in \mathbb{R}$ is a multiplicator, and $M$ is a $H \times H$ symmetric matrix of multiplicators.
$$d(tr(U'WUM)) = tr(dU'WUM) + tr(U'WdUM)$$
$$= tr(M'U'WdU) + tr(MU'WdU) = tr((2WUM)'dU)$$
$$\Leftrightarrow \frac{\partial tr(U'WUM)}{\partial U} = 2WUM$$

So, the first order conditions on $L$ yield:
$$\frac{\partial L}{\partial \mu} = 0 \Leftrightarrow \lambda'\lambda = 1 \quad (d)$$
$$\frac{\partial L}{\partial M} = 0 \Leftrightarrow U'WU = I_H \quad (e)$$
$$\frac{\partial L}{\partial \lambda} = 0 \Leftrightarrow \gamma(\lambda, U) = \mu\lambda \quad (f)$$
$$\frac{\partial L}{\partial U} = 0 \Leftrightarrow \Gamma(\lambda, U) = WUM \quad (g)$$

(d) and (f) imply that:
$$\mu = \frac{1}{\|\gamma(\lambda, U)\|}$$

Besides, (g) $\Leftrightarrow U = W^{-1}\Gamma(\lambda, U) M^{-1}$ and (e) imply that:
$$M^{-1}\Gamma(\lambda, U)'W^{-1}\Gamma(\lambda, U)M^{-1} = I_H \quad (h)$$

We see that $M = (\Gamma(\lambda, U)'W^{-1}\Gamma(\lambda, U))^{\frac{1}{2}}$ is symmetric and satisfies (h).
We easily check that a point that satisfies:
$$\lambda = \frac{\gamma(\lambda, U)}{\|\gamma(\lambda, U)\|} \quad ; \quad U = W^{-1}\Gamma(\lambda, U)(\Gamma(\lambda, U)'W^{-1}\Gamma(\lambda, U))^{-\frac{1}{2}}$$

is a critical point of the program.

**d)** $\forall G \in \mathbb{R}^{n \times H} : V = W^{-1}G(G'W^{-1}G)^{-\frac{1}{2}} = \arg\max_{U'WU = I_H} tr(U'G)$

Firstly, $V$ is such that $V'WV = I_H$. Then:
$$tr(U'G) = tr(U'G(G'W^{-1}G)^{-\frac{1}{2}}(G'W^{-1}G)^{\frac{1}{2}}) = tr(U'WV(G'W^{-1}G)^{\frac{1}{2}})$$

Under constraint $U'WU = I_H$, this is maximum for $U = V$. Indeed:
$$(G'W^{-1}G)^{\frac{1}{2}} = S\Delta S' \text{ with } \Delta = diag(\delta_h)_{h=1,\ldots,H}; \delta_h \geq 0 \ \forall \ h \text{ and } S \text{ orthogonal}$$

So, with $A = U'WV$:

$$tr(U'WV(G'W^{-1}G)^{\frac{1}{2}}) = tr(U'WV S \Delta S') = tr(\Delta S'AS) = \sum_{h=1}^{H} \delta_h (S'AS)_{hh} \quad (i)$$

With $S = [s_1, \ldots, s_H]$, we have:

$$(S'AS)_{hh} = \langle s_h | A s_h \rangle \quad (j)$$

Now $\forall s \in \mathbb{R}^h$:

$$\|As\|^2 = s'V'WUU'WVs = (Vs)'W \Pi_U Vs = \|Vs\|_W^2 \cos_W^2(Vs, \langle U \rangle) \leq \|Vs\|_W^2 \text{, and}$$

$$\|Vs\|_W^2 = s'V'WVs = s's \text{ , since } V'WV = I_H$$

which proves that:

$$\forall s \in \mathbb{R}^h, \|As\|^2 \leq \|s\|^2 \quad (k)$$

(j) and (k) imply that

$$(S'AS)_{hh} \leq 1 \quad (l)$$

This, in turn, in view of (i) and the positivity of the $\delta_h$'s, proves that:

$$tr(U'WV(G'W^{-1}G)^{\frac{1}{2}}) \leq \sum_{h=1}^{H} \delta_h = tr((G'W^{-1}G)^{\frac{1}{2}}) = tr(V'WV(G'W^{-1}G)^{\frac{1}{2}})$$